# A Delay Aware Routing Protocol for Wireless Sensor Networks

**Bhaskar Bhuyan**[1], **Nityananda Sarma**[2]

[1] **Dept of IT, Sikkim Manipal Institute of Technology**
**Mazitar, Rangpo, Sikkm-737136, INDIA**

[2] **Dept of Computer Science and Engineering, Tezpur University,**
**Napaam -784028, Assam, INDIA**

**Abstract**

Wireless Sensor Networks (WSNs) consist of sensor nodes which can be deployed for various operations such as agriculture and environmental sensing, wild life monitoring, health care, military surveillance, industrial control, home automation, security etc. Quality of Service (QoS) is an important issue in wireless sensor networks (WSNs) and providing QoS support in WSNs is an emerging area of research. Due to resource constraints nature of sensor networks like processing power, memory, bandwidth, energy etc. providing QoS support in WSNs is a challenging task. Delay is an important QoS parameter for forwarding data in a time constraint WSNs environment. In this paper we propose a delay aware routing protocol for transmission of time critical event information to the Sink of WSNs. The performance of the proposed protocol is evaluated by NS2 simulations under different scenarios.

## 1. Introduction

In recent years, wireless sensor network (WSNs) has become one of the cutting edge technologies for low power wireless communication. The fast development of low power wireless communication devices, the significant development of distributed signal processing, adhoc network protocols and pervasive computing have collectively set a new vision for wireless sensor network [1][2]. In majority of WSNs applications, a large number of sensor nodes are deployed to gather data based on application domains. This data collection process can be continuous, event driven and query based [3]. WSNs can be deployed in various domains and applications such as agriculture and environmental sensing, wild life monitoring, health care, military surveillance, industrial control, home automation, security etc. A simple model of wireless sensor networks is shown in Fig-1.

.

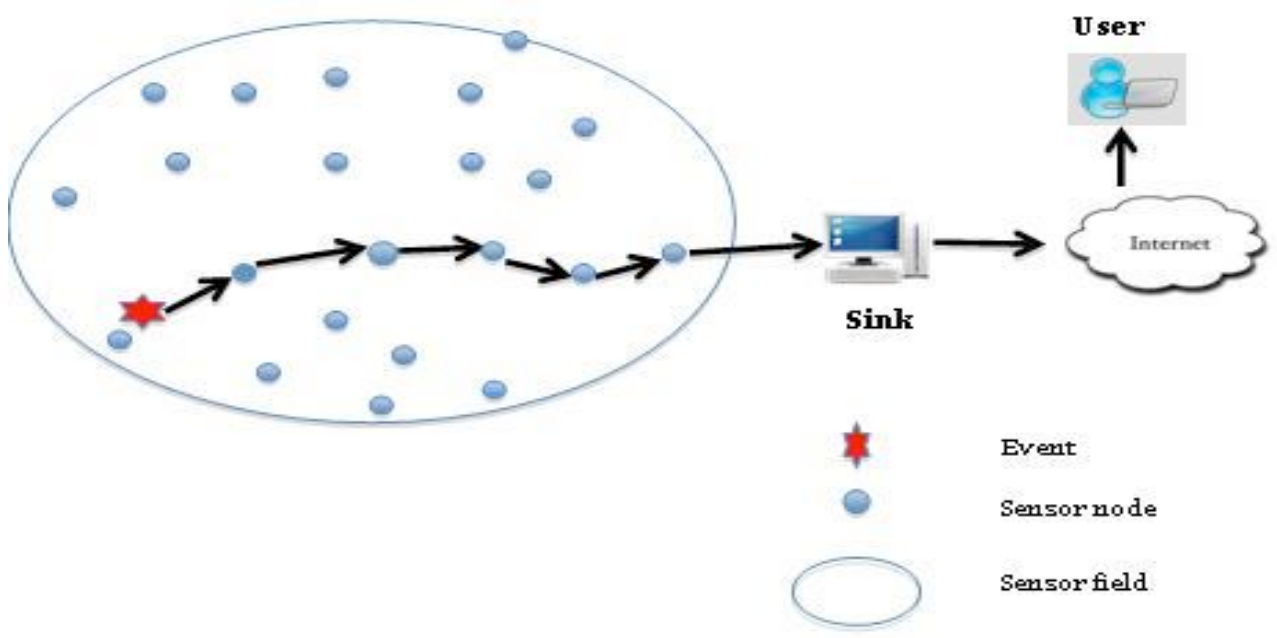

Fig-1. A simplified model for Wireless Sensor Networks

Provisioning of Quality of Service (QoS) is a challenging problem in WSNs. A simplified model for QoS in WSNs is shown in Fig-2 which is redrawn from [3]. In WSNs, delivery of certain time critical events to the Sink or Base Station within a specific deadline is an important aspect of QoS for the success of certain delay aware applications such as military surveillance, industrial monitoring, healthcare, disaster management etc. End to End Delay is considered to be one of the important parameter like the other QoS parameters such as reliability, energy, data accuracy, coverage etc. in Wireless Sensor Networks.

In this paper we propose a delay aware routing protocol for wireless sensor networks. The proposed protocol is a routing solution for the transmission of time critical event information to the Base Station within a deadline. It also ensures reliability by following multipath routing so that at least one copy of the event information is received by the Sink or Base Stations. We will use the term Sink or Base Stations interchangeably in this literature. The proposed protocol is designed for the application of typical WSNs where certain time critical events required to be detected and to be informed to the Sink within a certain delay limit with reliability.

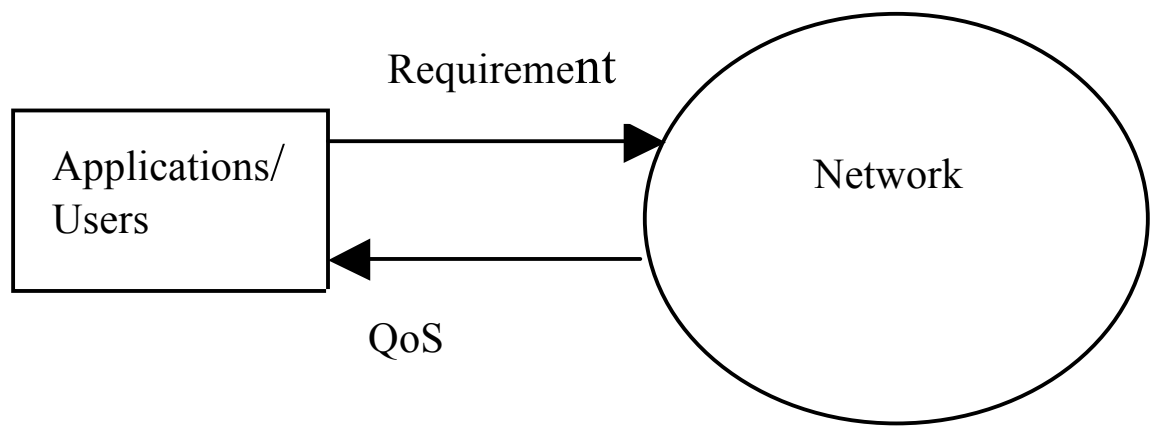

Fig-2. A simplified model for QoS in Wireless Sensor Networks

The remainder of this paper is organized as follows. In section 2, we present a review of the related work. In section 3, we have presented the detail working principle of the proposed protocol. In section 4, we have discussed the performance of the proposed protocol and finally in section 5 conclusions have been presented.

## 2. **Related Works**

QoS provision in wireless sensor networks is an emerging field of research and various literatures are available on this field. In this section we are briefly reviewing some of the QoS aware routing protocols for wireless sensor networks. Our survey mainly deals with delay and energy aware reliable routing protocols in wireless sensor networks.

T.He et. al proposed stateless real-time communication protocol called SPEED [7] in WSNs. It meets the end to end delay by enforcing uniform communication speed in every hop in the network through feedback control mechanism and non-deterministic QoS aware geographic-forwarding.

E.Felemban et al. proposed MMSPEED [9] which was designed as an extension of the SPEED to support multiple communication speeds over multi path and provides differentiated reliability.

Chenyang Lu et al. proposed a Real-time Architecture and Protocols (RAP) based on velocity [10]. RAP was designed to provide service differentiation by velocity-monotonic classification of packets in timeliness domain. The required velocity is calculated based on destination and packet deadline. Packets priority is assigned in the velocity-monotonic order so that a high velocity packet can be forwarded earlier than a lower velocity one.

Akkaya et al. proposed an energy aware QoS routing protocol [8] to find energy efficient paths along which end to end delay requirement can be fulfilled. Each node classifies the incoming packets and forwards the real time and non real time packets to different priority queues. The delay requirement is converted into bandwidth requirement. The use of class based priority queuing is complicated mechanism and costly for resource constraint sensors.

Chipara et al. proposed a Real time Power-aware Routing (RPAR) in [11]. This protocol was designed to achieve application specific communication delay with low energy cost by dynamic adjustment of transmission power routing decisions. The network topology changes frequently in this protocol due to dynamic adjustment of transmission power.

X.Hunag and Y.Fang have proposed Multi Constrained QoS Multi Path (MCMP) routing protocol [12] based on certain QoS requirements such as delay and reliability . In this protocol the authors have formulated an optimization problem for end to end delay and then, this problem is solved by linear programming. The protocol utilizes multiple paths to transfer packets with moderate energy expenditure. To fulfill the QoS requirement the protocol prefers to route information through paths having minimum number of hops which leads to more energy consumption.

Z.Lei et al. proposed FT-SPEED [13] which is an extension of SPEED protocol. This protocol takes care of routing voids while transmitting packets to its destination. Due to routing voids path length becomes longer and as a result of this data packets may not be delivered to its sink node before its deadline.

## 3. Proposed Delay Aware Routing Protocol

In this section a delay aware routing protocol is presented for wireless sensor networks. The protocol tries to transmit data packets to its base station or sink within the estimated deadline. The protocol constructs and maintains forwarding table based on the information gathered from its neighboring nodes. We have made the following assumptions while designing this protocol-

- Homogeneous sensor nodes are densely deployed in a sensor field.
- Each sensor node knows it's location by some localization techniques [15].
- The base station(s) or sink node(s) is/are GPS enabled and broadcast its location information to all sensor nodes.
- Sensor nodes are static with minimum node mobility.

- Radio range and initial energy level of all sensor nodes are equal

The main components of the proposed protocol are Neighborhood Management and Data Forwarding

## 3.1 Neighborhood Management

Each sensor nodes explore its one hop neighbors by the exchange of HELLO and ACK control packets. The fields of the Hello packets are as follows-

- source node id
- source node position
- distance to Sink

In response to Hello packet each node sends an ACK packet containing the following fields:

- Neighbor node id
- Neighbor node position
- Distance to Sink
- Residual Energy

The format of the HELLO packet and ACK control packets are shown in Fig-3(a) and (b). This neighbor discovery information is stored in each sensor node in the form of a table and updated periodically.

| Source node ID | Source node position | Distance to sink |
|---|---|---|

Fig-3(a). HELLO packet

| Neighbor node ID | Neighbor node position | Distance to sink | Residual energy |
|---|---|---|---|

Fig-3(b). ACK packet

Once the neighbor discovery phase is over each sensor node calculates the link delay to its neighboring nodes by broadcasting an echo packet and records the round trip time of the echo packet. For a pair of node (*ni, nj*) the link delay is calculated as shown in equation Eq.(1)

$$\text{link delay}_i^j = \text{Round trip time}/2 = (\text{Delay}_{MAC} + \text{Delay}_{queue} + \text{Delay}_{TX}) \times C_i^j \quad (1)$$

Where:

$\text{Delay}_{MAC}$: Channel access delay
$\text{Delay}_{queue}$: Queuing delay
$\text{Delay}_{Tx}$: Transmission delay and queuing delay.
$C_i^j$ : Transmission count

This delay may vary depending on the traffic load in the link. Based on the neighbor discovery information and estimated link delay, each sensor node maintains a data forwarding table which is looked up for routing data packets to its destination. The data forwarding table contains the following headings-

- Neighbor node id
- Neighbor node Position
- Distance to sink node
- Link delay
- Energy level

## 3.2 Packet Forwarding

When a sensor node receives some data due to occurrence of an event or obtains a data packets form its neighboring node, it forwards the data packet to the sink node through multi hop communication. For this each sensor node looks up it's routing table and forwards the data packet to the potential next hop towards the sink node. The deadline may be specified by the user or estimated by the source node of the event area. The data from sensor nodes is transmitted in the form a packet having the following control fields:

- source node ID
- sink node ID
- deadline

On receiving this packet the sensor node selects the next node for forwarding the packet based on the following conditions:

- The next hop selected should be closer to the destination w.r.t current node.
- Propagation speed of data packets provided on the selected link should meet the required propagation speed of on that link.

The required propagation speed depends on the estimated deadline. For node $n_i$, let $n_j$ is a neighbor of node $n_i$, where $i \neq j$. Let $d(n_i, n_j)$ represents the distance between node $n_i$ and $n_j$. The source node, *S,* calculates the required end-to-end propagation speed of data $V_{req}$ for the estimated deadline $t_{set}$ towards the sink node, *T*, as $V_{req} = d(S,T)/t_{set}$. A neighbor node will be selected as a forwarding node if the propagation speed, $V_{prov}$, on that link is greater than or equal to $V_{req}$ and the node is closer to the sink node than the current node. In every intermediate node between source and the sink the required and provided propagation speed are computed as follows:

Suppose,

$t_{set}$: Estimated deadline for the data packets at source node.

$t_l$: Time left to meet deadline. At the source node $t_l$, is equal to $t_{set}$ .

In every intermediate node $t_l$ is updated as $t_l = t_l$ – link delay. The required speed is recalculated as shown in Eq.(2)

$$V_{req} = d(n_i, T)/(t_l - \text{link delay}) \quad (2)$$

Here, $n_i$ is an intermediate node in a path between source and the sink. When i = 0, $n_i$ is the source node S.

Similarly, propagation speed provided on each selected link is computed as shown in Eq. (3)

$$V_{prov} = d(n_i, T) - d(n_{i+1}, T)/\text{link delay}_i^{i+1} \quad (3)$$

A neighbor node will be selected as a forwarding node if the provided speed, $V_{prov}$, is greater than or equal to required speed, $V_{req}$, and the forwarding node is closer to the destination w.r.t current node.

A copy of the data packet is also forwarded to an alternative neighbor node if the alternative link fulfills the required propagation speed. This will be helpful to forward data packets if the primary link fails. This redundancy will also help in ensuring reliability. This data duplication is done at the source node only, which senses the occurrence of an event. All intermediate nodes forward the packet only along single route to the destination.

# 4. Performance Evaluation

## 4.1 Simulation Environment

The proposed protocol has been simulated using network simulator ns-2.35. Nodes are deployed in a flat grid area of 600x400 square meters. We have simulated our protocol by varying the no of nodes as 50, 75, 100, 125, 150. The simulation time is also varied as 100s, 200s, 300s, 400s, 500s. We have used IEEE802.11 MAC protocol for our experiments. The various simulation parameters chosen are summarized in the Table-1 below.

Table-1: Simulation Environment

| | |
|---|---|
| No of Nodes | 50,75,100,125,150 |
| Simulation Time | 100s, 200s, 300s, 400s, 500s |
| Node Placement | Flat Grid |
| Area | 600x400 |
| MAC protocol | 802.11 |
| Propagation model | Two ray ground |
| Transmission Range | 250m |
| Traffic Type | CBR |

## 4.2 Simulation Results

The main performance metrics which are evaluated in our experiments are (i) end to end delay (ii) packet delivery ratio (iii) deadline miss ratio

### 4.2.1 End to End Delay

The end to end delay is the average end to end delay for all successfully received packets. We have evaluated average end to end delay for five different set ups by varying the number of nodes as 50, 75,100,125 and 150. For each set up we have varied the simulation time as 100s, 200s, 300s, 400s and 500s. In Fig-4, the average end to end delay for different number of deployed nodes with respect to simulation time is shown.

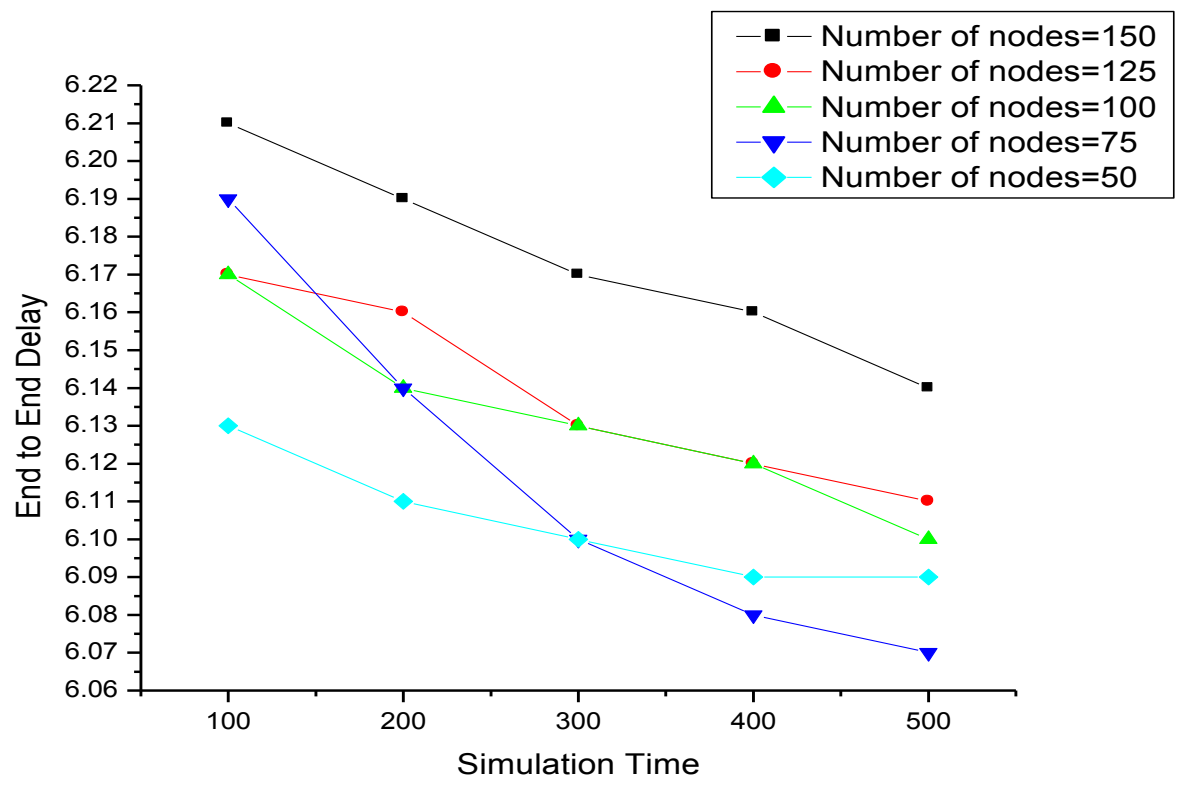

Fig-4. Average End to End Delay when simulation time is varied

The figure shows that average end to end delay decreases with the increase of simulation time.

### 4.2.2 Packet Delivery Ratio

The packet delivery ratio is the ratio of total number of packets received to the number of packet sent without considering the deadline. We have evaluated packet delivery ratio for reliability purpose. Here also we have evaluated this performance metric for five different set ups by varying the number of deployed nodes as 50, 75,100,125 and 150. For each set up we have varied the simulation time as 100s, 200s, 300s, 400s and 500s as mentioned earlier. Fig-5 shows the packet delivery ratio for different number of nodes with respect to varying simulation time.

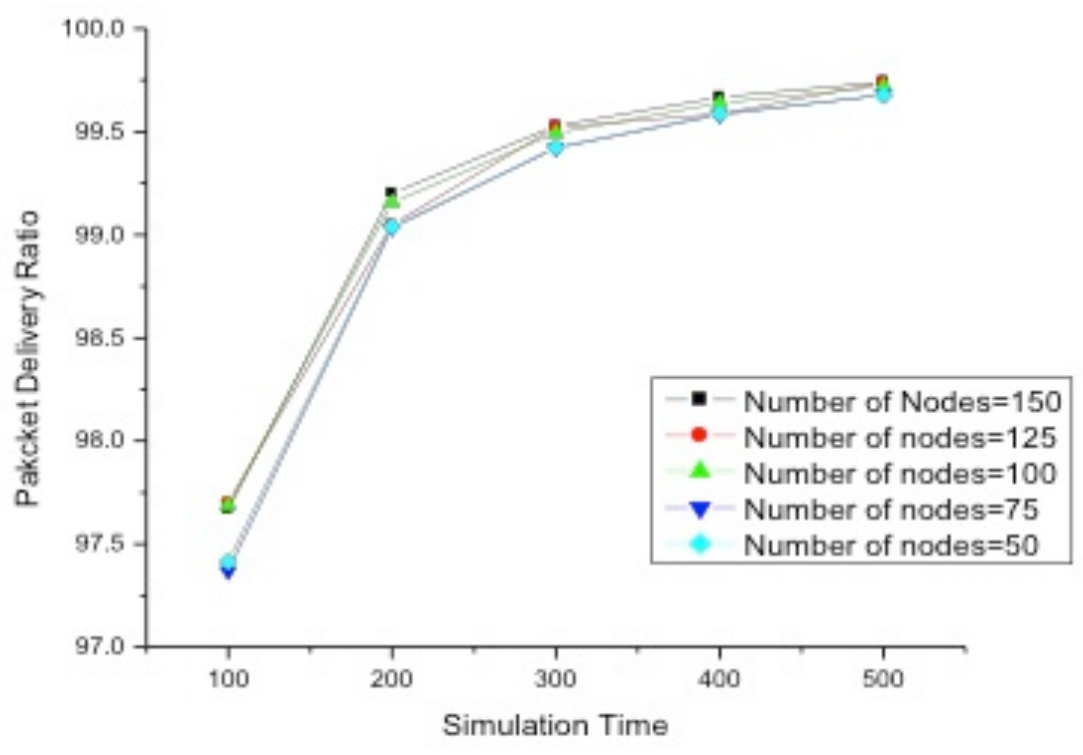

Fig-5. Packet delivery ratio when simulation time is varied for different number of nodes.

Fig-5 shows that packet delivery ratio increases with increases of simulation time. This is clear from the fact that with increase in simulation time more packets will be generated by sources and effectively packet delivery ratio will be increased.

### 4.2.3 Dead Line Miss ratio

Deadline miss ratio is the fraction of packets that missed their deadlines. This is computed to evaluate the performance of the packet forwarding policy. Fig-6 shows the deadline miss ratio of the proposed protocol with respect to varying number of sensor nodes. Here the simulation time is set to 500 sec and deadline is set to 6 msec. Each source sends a packet in 1s interval.

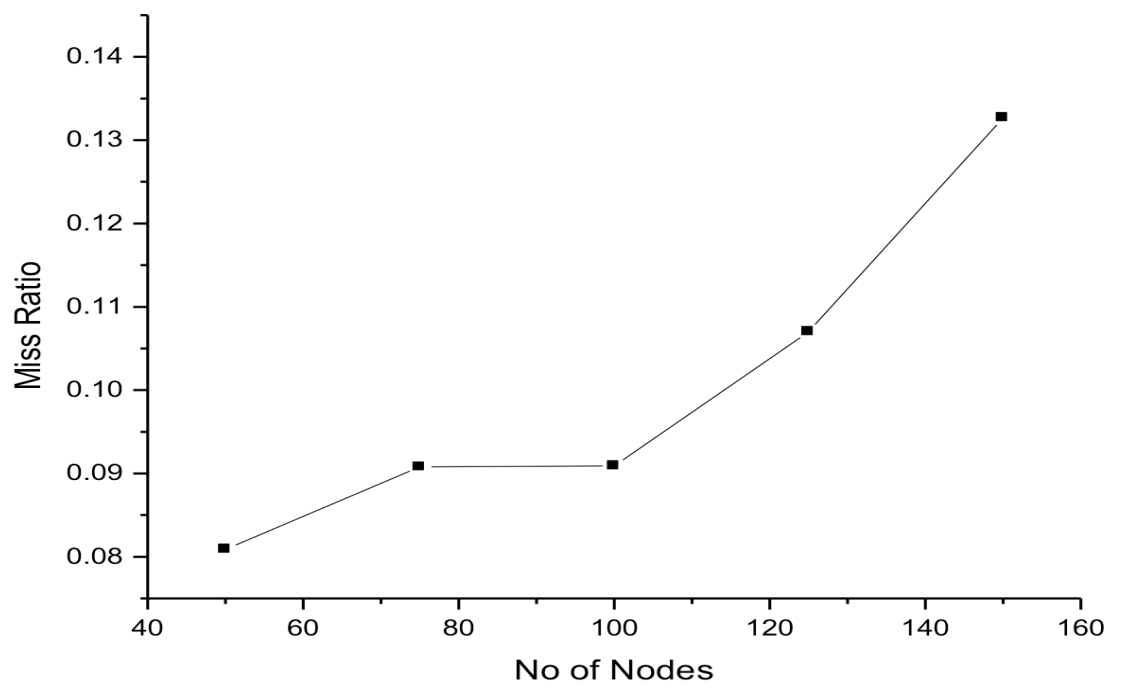

Fig -6. Dead line miss ratio when number of nodes are varied

From the graph shown in Fig-6, it can be observed that more packets miss deadline as the number of nodes are increased.

Next we have evaluated the deadline miss ratio by varying the deadlines. Deadlines are varied as 6ms, 7ms, 8ms, 9ms and 10ms. We have performed this experiment for five different set ups by varying the number of nodes as 50, 75, 100, 125 and 150. The simulation time is set to 500s. Each source nodes sends packets to sink in 1s interval. Fig-7 shows the deadline miss ratio for different number of nodes with respect to varying deadlines. It is obvious from the graph that deadline miss ratio decreases with the increase of deadlines irrespective of any number of nodes deployed as shown in the Fig-7.

We have also evaluated the performance of our forwarding policy by changing the load at the source. Here we have evaluated deadline miss ratio by varying the packet generation interval between two consecutive packets at the source as 1s, 2s, 3s, 4s and 5s. We have run this experiment for three different set ups by varying the node numbers as 50,100 and 150. Fig-8 shows deadline miss ratio for node number 50,100 and 150 by varying the packet interval time at the sources.

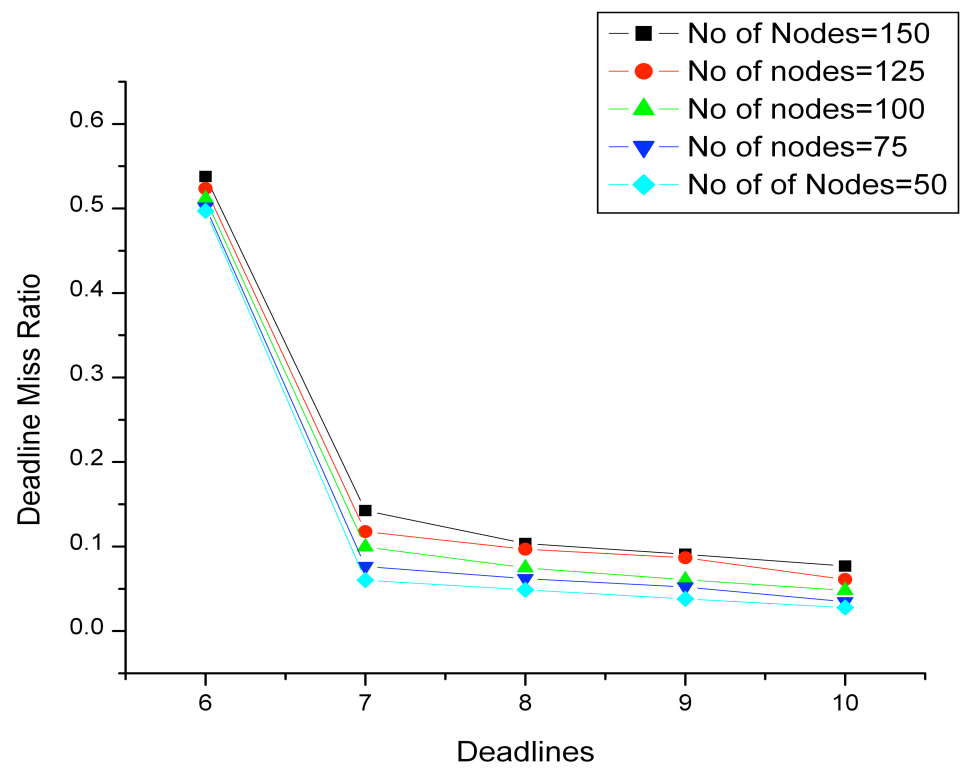

Fig-7. Deadline Miss Ratio varying the Deadlines

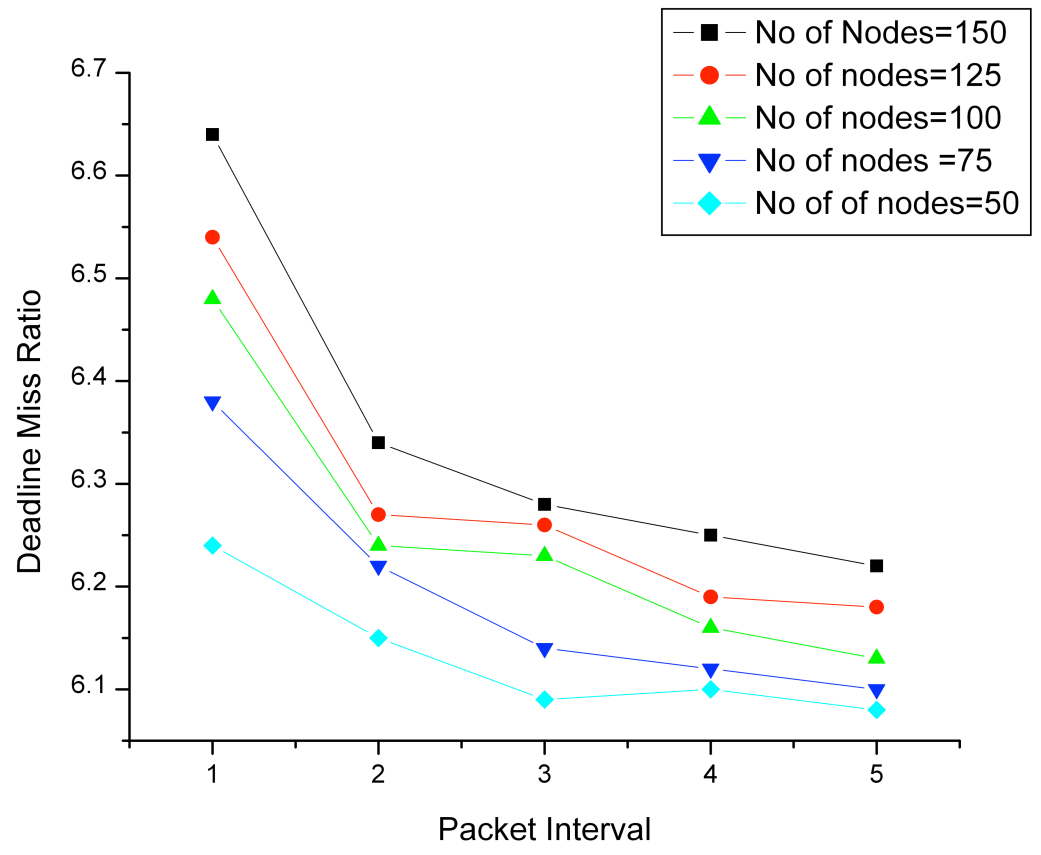

Fig-8. Deadline Miss Ratio varying Packet Interval

The graph in Fig-8 shows the fall of deadline miss ratio as we reduce the load at the sources by increasing the packet interval time between the consecutive packets.

## 5. Conclusions

In this paper we present a delay aware routing protocol for wireless sensor networks. We have simulated the protocol in ns-2.35 environment by considering maximum of 150 nodes deployed in a flat area of 600X400 square meter. The performance of the proposed protocol is evaluated by measuring the end to end delay and packet delivery ratio varying the number of deployed nodes and simulation time. We have also evaluated the effectiveness of the proposed protocol by measuring the deadline miss ratio varying the estimated deadlines for data delivery and load in the network. A comparative simulation analysis of the proposed protocol with its related protocols is under progress as a part of its future work.

**Bhaskar Bhuyan** is an Associate Professor in Sikkim Manipal Institute of Technology under Sikkim Manipal University, Sikkim, India. .He received his Bachelor of Engineering degree in Computer Science and Engineering from Motilal Nehru Rehional Engineering College(now NIT) , Allahabad , India. He completed M.Tech in information Technology from Tezpur University, Assam, India. Currently he is pursuing PhD in Computer Science and Engineering from Tezpur University. His research interest includes Wireless Sensor Networks, Mobile Adhoc Networks and Cloud Computing. He is a member of IEEE and System Society of India.

**Nityananda Sarma** is a Professor in the Department of Computer Science & Engineering at Tezpur University, Assam, India. He received his Bachelor of Engineering degree in Computer Science & Engineering from Jorhat Engineering College, Assam, India. He did Master of Technology in Computer Science & Engineering and Ph.D in Computer Science & Engineering from Indian Institute of Technology Kharagpur and Indian Institute of Technology Guwahati respectively. His areas of research include - Ad Hoc Networks, QoS supports in Wireless Networks, Channel Assignment in Cognitive Radio Networs and WDM Optical Networks He is a Professional Member of IEEE, ACM and Fellow of IETE.